# Nonradiative to Superscattering Switch with Phase-Change Materials

Sergey Lepeshov[1*], Alex Krasnok[2*], and Andrea Alù[2]

[1]ITMO University, St. Petersburg 197101, Russia

[2]Photonics Initiative, Advanced Science Research Center, City University of New York, New York 10031, USA

*s.lepeshov@gmail.com*, *aalu@gc.cuny.edu*

*These Authors contributed equally

*Phase-change materials (PCMs) can switch between different crystalline states as a function of an external bias, offering a pronounced change of their dielectric function. In order to take full advantage of these features for active photonics and information storage, stand-alone PCMs are not sufficient, since the phase transition requires strong pump fields. Here, we explore hybrid metal-semiconductor core-shell nanoantennas loaded with PCMs, enabling a drastic switch in scattering features as the load changes its phase. Large scattering, beyond the limits of small resonant particles, is achieved by spectrally matching different Mie resonances, while scattering cancellation and cloaking is achieved with out-of-phase electric dipole oscillations in the PCM shell and Ag core. We show that tuning the PCM crystallinity we can largely vary total (~15 times) and forward (~100 times) scattering. Remarkably, a substantial reconfiguration of the scattering pattern from Kerker (zero backward) to anti-Kerker (almost zero forward) regimes with little change (~5%) in crystallinity is predicted, which makes this structure promising low-intensity nonlinear photonics.*

Scattering of electromagnetic waves lies at the heart of most experimental techniques across the entire spectrum, and hence it is of great interest for modern science and technologies. The interaction of light with individual nanoscale objects is therefore vitally important for optics and nanophotonics, with practical significance for various applications, including sensing [1,2], imaging [3,4], cloaking [5,6], and functional devices [7,8]. Progress in nanofabrication techniques in the last few decades has led to the fabrication of optical nanostructures with anomalous light scattering features. Examples of such effects include *cloaking-like nonradiative* states [9,10] and *superscattering* resonant states [11,12], beyond the conventional limits predicted for subwavelength passive objects. Reconfigurable scattering systems comprising both these anomalous scattering features pave the way to tunable nanophotonic devices with highly nonlinear and fast responses.

According to Mie theory, the total scattering cross section (SCS) of a finite object can be expressed as a weighted sum of the scattering contributions of different orthogonal spherical harmonics, defined outside the smallest sphere enclosing the three-dimensional scatterer. For a spherically-symmetric object of radius $R$, the SCS can be written as [13]

$$SCS = \frac{\lambda^2}{2\pi} \sum_{l=1}^{N} (2l+1) \left( |c_l^{TM}|^2 + |c_l^{TE}|^2 \right), \quad (1)$$

where $l$ defines the order of the scattering channel and equals to its total angular momentum, $\lambda$ is the wavelength in the surrounding medium, $N \approx kR$, $k = 2\pi/\lambda$, and $c_l^{TM}$ and $c_l^{TE}$ are the scattering coefficients[13]. For a spherical scatterer, these coefficients can be written in the form $c_l^i = -U_l^i / (U_l^i - jV_l^i)$ [14], where $i$ stands for TE or TM polarization and the quantities $U_l^i$ and $V_l^i$ are expressed as a combination of spherical Bessel and Neumann functions[15]. The scattering coefficients indicate the contribution of the different scattering channels, and the superscript TM or TE indicates whether the vector spherical harmonic have electric or magnetic field orthogonal to the radial direction.

In recent years, it has been observed that SCS may be made arbitrarily small in a desired region of the spectrum with crafted engineering of the scatterer. In this regime, the object does not scatter light to any channel and appears invisible to an external observer. For large-size objects ($kR \gg 1$) this requires making all relevant scattering amplitudes zero ($U_l^i = 0$), which gives rise to the cloaking effect. Nowadays different techniques exist to achieve it, including plasmonic and mantle cloaking[16,17], transformation-optics cloaking[18,19], transmission-line cloaking[20], among many other techniques, and we refer an interested reader to topical reviews (e.g.,[21]) for more detailed discussions on this broad area of research.

For a subwavelength object ($kR \leq 1$), the SCS is typically dominated by the dipole term ($l = 1$), so that only $c_1^i = -U_1^i / (U_1^i - jV_1^i)$ governs the scattering properties. A zero of its amplitude, i.e., $U_1 = 0$, gives rise to cloaking, for which the scattering becomes remarkably smaller than usual. In turn, $V_1^i = 0$ defines the scattering resonances of the system. If the two conditions are met for closely spaced frequencies, at the scattering zero we achieve an *anapole state*[22], for which the internal electric field is very large, despite the scattering being low. In fact, absence of radiation loss can further enhance the resonant fields in the scatterer, boosting nonlinear effects, such as third harmonic generation and four-wave mixing[23–26].

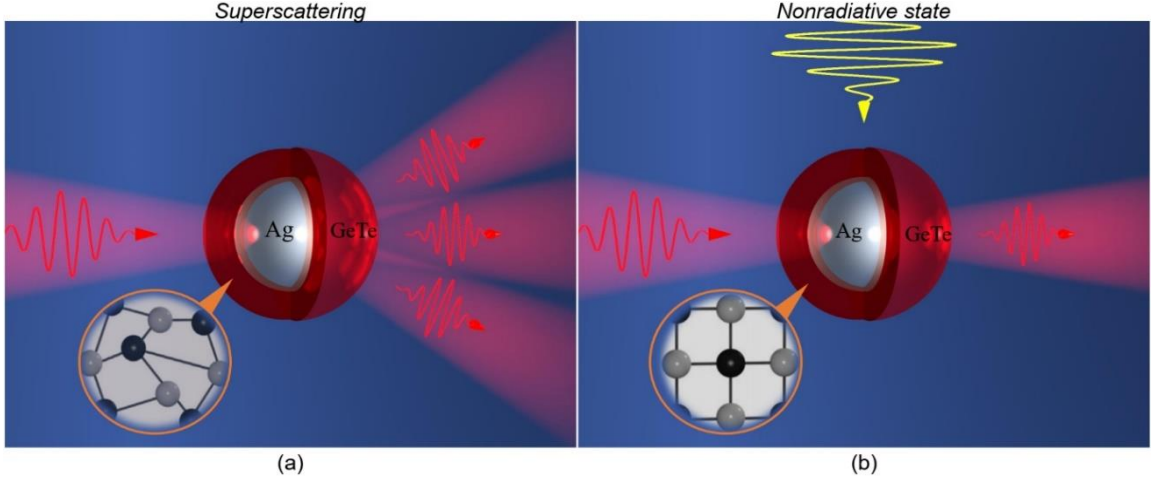

**Figure 1.** Schematic representation of a hybrid nanoantenna consisting of silver (Ag) core and a phase-change material (GeTe) shell. The antenna is designed to support superscattering in its amorphous phase (a) and a nonradiative state in the crystalline phase (b), with strong tunability of the scattering properties. Phase changing can be implemented by the same signal beam (red pulses) or with an additional laser beam (yellow pulse in panel (b)).

Energy conservation in passive systems imposes a limit on the total scattered energy into one scattering channel. As a result, the maximum SCS contribution from any given harmonic is $(2l+1)\lambda^2/2\pi$, achievable only when $V_l^i = 0$. For the dipole resonance ($l=1$), it yields $3\lambda^2/2\pi$ [27]. It has been recently demonstrated that the SCS of a small object can exceed this single-channel limit by employing the excitation of multipole resonances (at least two) at the same frequency. This superscattering regime has been proposed in Ref. [11] and theoretically proposed for scatterers of various geometries, including core-shell cylinders[11,28–30], core-shell spheres[31–33], double-slit structures[34], nanodisks[35] and experimentally demonstrated at microwaves[12].

Thus, nonradiative states (anapole, cloaking) and superscattering regimes present two *opposite extreme scattering phenomena*, and their implementation in one single reconfigurable scattering system opens a pathway to highly tunable optical systems[34].

In this paper, we propose a hybrid nanoantenna comprising a phase-change material (GeTe) possessing superscattering and nonradiative regimes in different crystalline phases of GeTe. The nanoantenna consists of a silver (Ag) core and GeTe shell and is schematically presented in **Figure 1**. The core and shell have subwavelength radii $R_{core}$ and $R_{shell}$, respectively. As demonstrated below, the Ag particle provides strong electric field enhancement at the plasmonic dipole (ED) resonance, whereas the GeTe shell is governed by the magnetic dipole (MD)[2,36–39] Mie resonance (and nonresonant contribution of ED). GeTe belongs to the class of PCMs possessing the ability of rapidly switching between different crystalline phases with a substantial change in dielectric permittivity[40]. Note that we use GeTe material in its transparency window (0.4 – 1 eV) where its extinction coefficient is smallest[40].

The nanoparticle is designed to support a superscattering regime in the amorphous phase [schematically shown in **Figure 1**(a)] and a nonradiative state in the crystalline phase [**Figure 1**(b)] offering extreme tunability of the scattering properties. Namely, tuning of the PCM crystallinity leads to a tremendous change in the total (~15 times) and forward (~100 times) scattering. We also observe a substantial reconfiguration of the scattering pattern from Kerker (zero backward) to anti-Kerker (almost zero forward) with a tiny change in crystallinity. Practically, the changing of phase can be implemented by the same signal beam (red pulses) or with an additional laser beam, as shown by the yellow pulse in **Figure 1**(b).

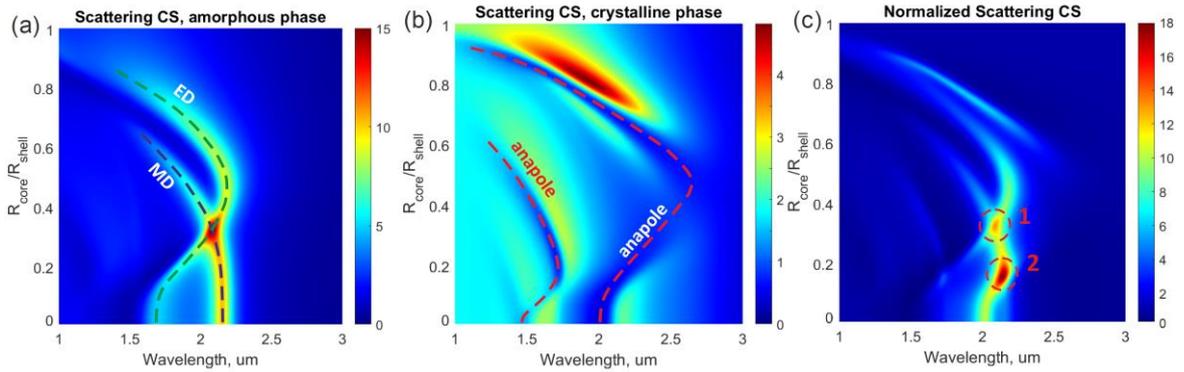

**Figure 2.** SCS of the hybrid Ag-GeTe core-shell nanoantenna versus $R_{core}/R_{shell}$ ratio and the wavelength in (a) amorphous and (b) crystalline phase. The shell radius $R_{shell}$ is fixed to 270 nm. (c) Ratio of SCS of the antenna in the amorphous phase to its SCS in the crystalline phase. Regimes #1 and #2 correspond to transitions from superscattering to anapole and from MD resonance to anapole, respectively.

To begin with, we optimize the structure geometry to achieve the superscattering regime in the amorphous phase. Since the overall size is subwavelength, superscattering is achieved for SCS larger than $3\lambda^2/2\pi$. To find the optimal layout, we fix the shell radius $R_{shell}$ to 270 nm and vary the core radius. For analysis, we use Mie theory for multilayered spheres[14], with dielectric permittivities of Ag and GeTe taken from[40,41], respectively. The results of analytical calculations of SCS versus $R_{core}/R_{shell}$ ratio and wavelength are summarized in **Figure 2**(a). The ED and MD resonances may be adjusted at $R_{core} = 0.33R_{shell}$ with a strong increase in the total scattering, up to ~15 in units of the geometrical cross-section, $\pi R^2_{shell}$. In this regime the system beats the one channel limit, giving rise to superscattering.

While the phase of GeTe material changes, the scattering properties of the nanoantenna are fundamentally transformed. In the fully crystalline phase, the antenna possesses two zero scattering lines, depicted in **Figure 2**(b). The one occurring at shorter wavelengths corresponds to an anapole state or a zero of the ED scattering amplitude, similarly to the phenomenon recently reported in [42]. The second zero SCS line corresponds to a cloaking-like scattering dip[16]. This regime is caused by scattering cancellation of dipole moments induced in the core and shell and oscillating in opposite phase. Remarkably, the superscattering regime in the amorphous phase overlaps nicely with the cloaking regime in the crystalline phase, making this structure tunable between these two opposite scattering anomalies. The ratio of SCS in amorphous phase to its SCS in crystalline phase is presented in **Figure 2**(c), manifesting several possible scattering conditions when the amorphous-to-crystalline phase change leads to considerable scattering tuning. The regime #1 corresponds to the transition from superscattering to the anapole regime, while the regime #2 is caused by a switch between the MD resonance and the anapole state.

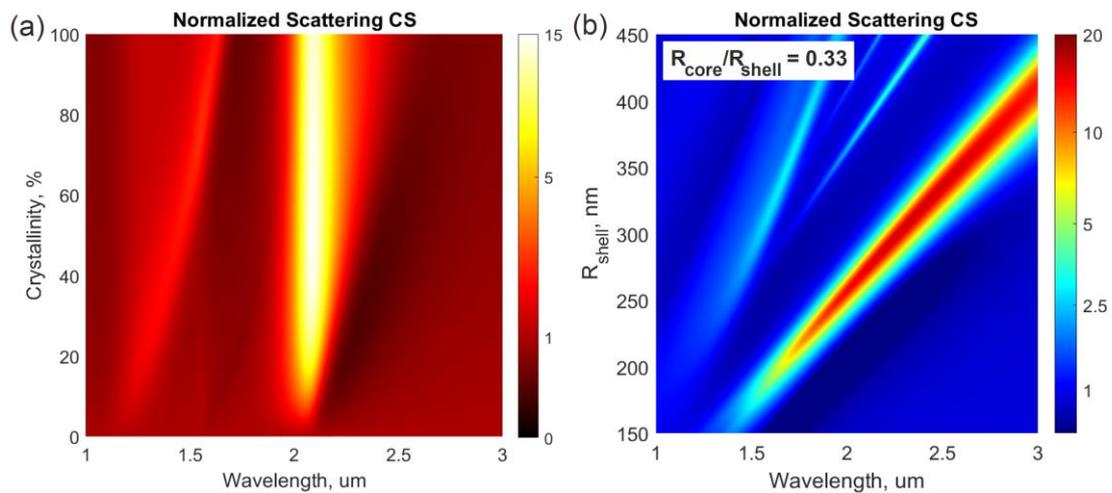

**Figure 3.** (a) SCS of the Ag-GeTe antenna with $R_{shell}$ = 270 nm and $R_{core}$ = 0.33$R_{shell}$ in the amorphous phase normalized to the SCS in the variable crystallinity phase versus crystallinity and the wavelength. (b) SCS of the Ag-GeTe antenna in amorphous phase normalized to its SCS in the 50% crystallinity phase versus radius $R_{shell}$ and the wavelength.

PCMs enable a smooth change between different crystalline states[43], although some of them may be unstable. Here, we explore the SCS dependence on the GeTe crystallinity and define the amount of crystallinity needed for optimal tuning. In our analysis, we vary the permittivity of the shell $\varepsilon_{shell}$ according to the formula

$$\varepsilon_{shell}(x) = x\varepsilon_{cr} + (1-x)\varepsilon_{am} \qquad (2)$$

where $\varepsilon_{cr}$ and $\varepsilon_{am}$ are permittivities of GeTe in crystalline and amorphous phases respectively, $x$ is the crystallinity. **Figure 3**(a) demonstrates the SCS of the Ag-GeTe nanoparticle in the amorphous phase normalized to the SCS in the variable crystallinity phase versus crystallinity and wavelength. The structure parameters are fixed to $R_{shell}$ = 270 nm and $R_{core}$ = 0.33$R_{shell}$, according to the results in **Figure 2**. One can observe a quick drop of SCS at wavelengths ~2 um with change from amorphous to 10% crystallinity reaching the maximum at ~50% crystallinity. The subsequent growth of crystallinity does not change the scattering ratio significantly.

Interestingly, an appropriate modification of the antenna geometry allows one to adjust the operation wavelength over a wide spectral range, as seen in **Figure 3**(b), where the SCS of the Ag-GeTe particle in the amorphous phase normalized to its SCS in the 50% crystallinity phase versus radius $R_{shell}$ and wavelength is shown. The ratio of radii is fixed to $R_{core}$ = 0.33$R_{shell}$. This design makes possible to tune the operation frequency over a wide bandwidth, including the telecom wavelengths around 1.5 um.

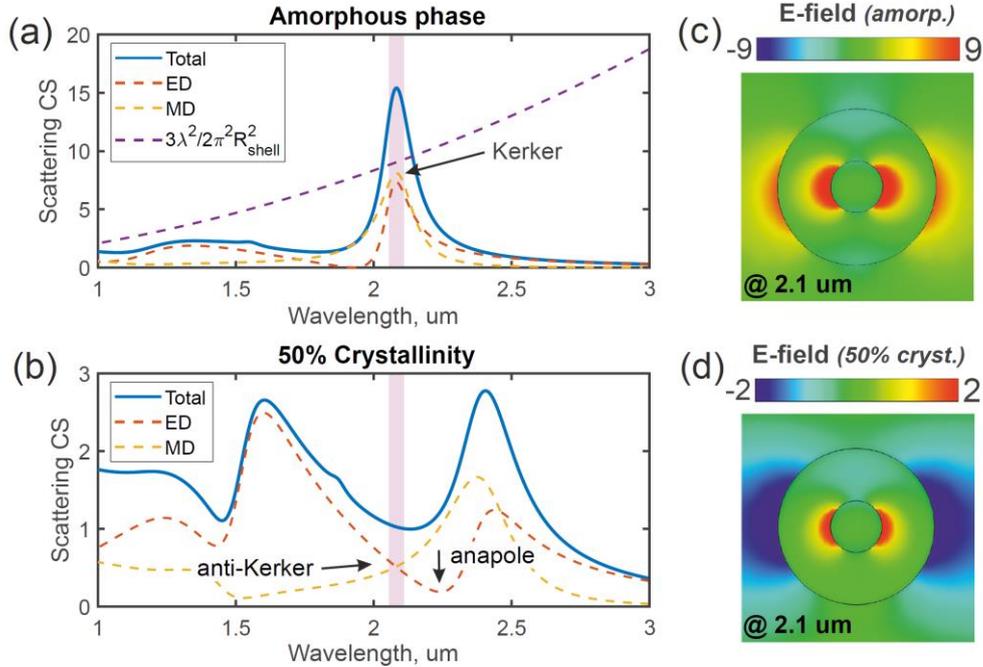

**Figure 4.** Scattering cross section of the hybrid nanoantenna with $R_{shell}$ = 270 nm and $R_{core}$ = 0.33$R_{shell}$ in (a) the amorphous phase and (b) the phase with 50% crystallinity. (c), (d) Electric

field enhancement distributions of the same core-shell in (c) the amorphous phase and (d) the phase with 50% crystallinity at the wavelength of 2.1 um.

A multipole decomposition analysis provides important insights into the scattering phenomenon [44,45]. The results of the multipole analysis for our structure in amorphous and 50% crystallinity phases is presented in **Figure 4**(a) and (b), respectively. We find that the spectral overlap of the ED and MD resonances at 2.1 µm, at which the total SCS (blue curve) beats the one-channel limit (pink dashed curve). Higher-order multipoles contribute weakly to the SCS and are not shown in this figure. The E-field distribution inside the system [**Figure 4**(c)], demonstrates in-phase oscillation of the polarization vector with 9 times enhancement. As the phase changes to 50% crystallinity, the ED and MD resonances are totally detuned from each other, so that the system turns into a nonresonant scattering regime. The anapole scattering regime at the slightly larger wavelength (~2.2 um) manifests itself as a dip in the ED scattering amplitude. Note that our structure operates slightly out of this regime (~2.1 um), where the ED and MD amplitudes are equal and have opposite phases [**Figure 4**(d)]. This out-of-phase ED and MD excitation with equal amplitudes supports the *anti-Kerker* scattering regime, in contrast to the Kerker regime, in which they have the same phase. In the Kerker regime a particle is known to scatter light forward, i.e., in the direction of the impinging wavevector with suppressed backward scattering, while in the anti-Kerker regime the structure scatters mostly in the backward direction with small forward scattering [46]. The optical theorem imposes a finite forward scattering for any passive system, but for small scatterers the scattering pattern can be largely tilted in the backward direction [47], as we find here.

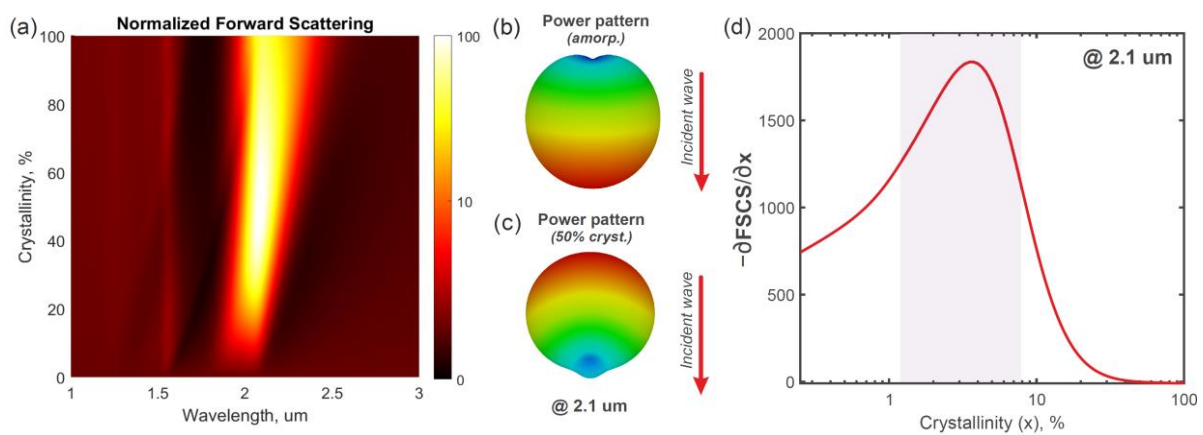

**Figure 5.** (a) Forward SCS of the hybrid nanoantenna ($R_{shell}$ = 270 nm and $R_{core}$ = 0.33$R_{shell}$) in the amorphous phase normalized to its FSCS for different crystallinity versus crystallinity and wavelength. (b),(c) Scattering power patterns of the antenna in the amorphous phase (b) and 50% phase (c) at 2.1 um. Red arrows show the wavevector direction of the impinging wave. (d) Derivative of FSCS by the crystallinity.

The Kerker regime observed here indicates that the structure possess stronger tunability in the forward scattering. In fact, **Figure 5**(a) demonstrates that the forward SCS (FSCS) of the hybrid nanoantenna ($R_{shell}$ = 270 nm and $R_{core}$ = 0.33$R_{shell}$) in the amorphous phase normalized to its FSCS in the 50% phase. It is seen that the FSCS reaches ~100 at 50% crystallinity and stays very large (50-100) in a wide range of crystallinity. Figure 5(b) and (c) demonstrate the antenna scattering power patterns in the amorphous phase and 50% crystallinity phase at the wavelength of 2.1 um, respectively. Here we observe the expected behavior of the scattering pattern, namely reconfiguration from zero backward to almost zero forward as the material crystallinity is changed.

Practically, it is important to define the crystallinity region where the tuning is the strongest. For this purpose, we calculate the derivative of FSCS by the crystallinity at 2.1 um, **Figure 5**(d). This result shows that in the most interesting scenario when one starts from the amorphous phase, the antenna is very tunable, reaching a tuning sensitivity maximum ~5%, where a little change in crystallinity leads to a strong change in antenna FSCS.

## Conclusions

We have proposed a reconfigurable hybrid metal-semiconductor core-shell nanoantenna made of silver (Ag) core and phase-changing GeTe material. The antenna demonstrates switching between superscattering and a nonradiative cloaking state, i.e., zero of scattering amplitude. The superscattering regime has been achieved by spectral matching of different Mie resonances, while the out-of-phase oscillation of electric dipoles in the PCM shell and Ag core gives rise to the cloaking state. We have shown that tuning of the PCM crystallinity leads to a large change in total (~15 times) and forward (~100 times) scattering. A drastic reconfiguration of the scattering pattern from Kerker (zero backward) to anti-Kerker (almost zero forward) with a tiny change in crystallinity has been observed. The proposed functionality is promising for low-intensity nanophotonic applications.